\documentclass{sig-alternate}

\usepackage{balance}
\usepackage{flushend}

\begin{document}
%

\title{Modeling Actuation Constraints for IoT Applications}
%
%
%
%
%

\numberofauthors{1} 
\author{
\alignauthor
Bharathan Balaji, Brad Campbell, Amit Levy, Xiaozhou Li,\\ Addison Mayberry, Nirupam Roy, Vasuki Narasimha Swamy, Longqi Yang, \\
Victor Bahl, Ranveer Chandra, Ratul Mahajan\\
        \vspace{5mm}
        \affaddr{Microsoft Research Student Summit -- Internet of Things Working Group}\\
}

\maketitle
\begin{abstract}
Internet of Things (IoT) promise to bring ease of monitoring, better efficiency and innovative services across many domains with connected devices around us. With information from critical parts of infrastructure and powerful cloud based data analytics, many applications can be developed to gain insights about IoT systems as well as transform their capabilities. Actuation applications form an essential part of these IoT systems, as they enable automation as well as fast low level decision making. However, modern IoT systems are designed for data acquisition, and actuation applications are implemented in an adhoc manner. We identify modeling constraints in a systematic manner as indispensable to support actuation applications because constraints encompass high level policies dictated by laws of physics, legal policies, user preferences. We explore data models for constraints in IoT system with the example of a home heating system, and illustrate the challenges in enforcing these constraints in the IoT system architecture.
\end{abstract}




\section{Introduction}
Internet of Things (IoT) promises to expand connectivity beyond computers and phones to physical entities embedded across various sectors: infrastructure, health, transportation, agriculture~\cite{evans2011internet}. Ubiquitous connectivity will provide us with a holistic view of the entire system, and insights can be drawn at various levels of decision making. Device to device communication can lead to unprecedented automation, and initial innovation from such connectivity can already be seen with Vehicle to Vehicle communication~\cite{reichardt2002cartalk} and multi-robot coordination~\cite{burgard2005coordinated}.

\begin{figure}[t!]
	\includegraphics[width=\linewidth]{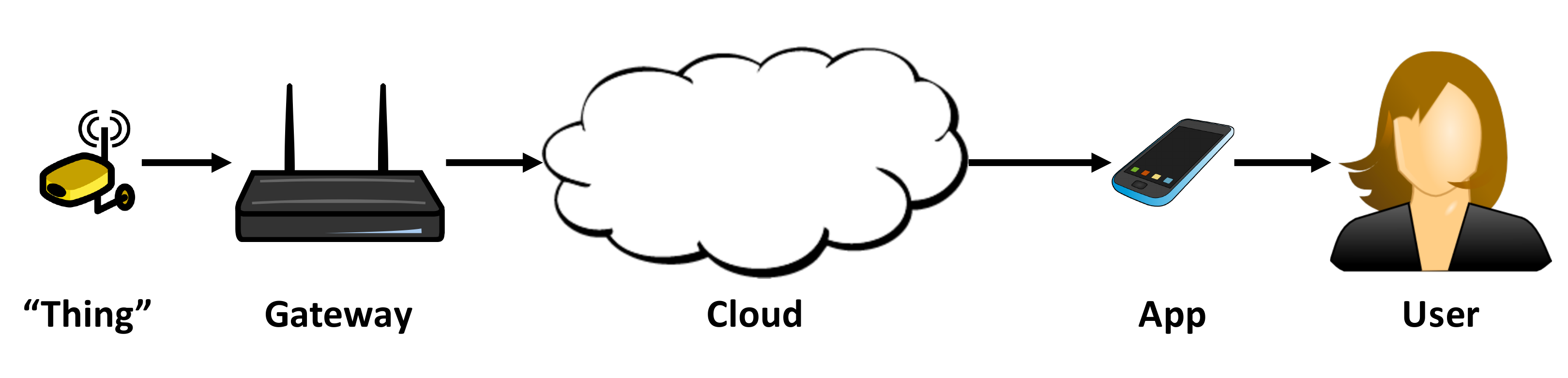}
	\caption{Typical Data Flow Architecture in IoT Systems}
	\label{fig:iot_pipeline}
\end{figure}

Figure \ref{fig:iot_pipeline} shows a typical pipeline in modern IoT systems. Significant progress has been made in each layer, from low power sensors to data analytics in the cloud. To process the data generated by IoT, however, we need a common data model across different sources. Standardized APIs exposed using RESTful web services have been proposed for abstracting away different communication protocols used by IoT vendors~\cite{dawson2010smap}. Domain specific schema and ontologies have been proposed to map the IoT data to a standard representation~\cite{barnaghi2012semantics,compton2012ssn}.

The data models proposed thus far can map metadata associated with the IoT device, such as measurement provenance, its accuracy, and domain specific attributes. With a semantic ontology, it is possible to represent relationship between system entities and incorporate these dependencies in the data analysis. However, much of the data modeling has been focused on representing information for data analysis. The data models provide little support for applications that can make actuation decisions within the IoT system, and mechanisms for control are implemented in an adhoc manner by application developers.

Control applications form an essential part of the IoT ecosystem as they enable automation and implementation of high level policies with little oversight. Examples include autonomous mining, irrigation systems, drone delivery system, public transportation. Each of these control applications need to operate within \emph{constraints} dictated by the domain. For example, an irrigation system needs to water the plantation based on soil type, plant type, ground slope and should not water sidewalks or humans when present. To implement these control applications, a developer is expected to acquire deep domain knowledge and incorporate constraints that encapsulate laws of physics, user preferences into the application. 

Just as modern computer operating systems provide memory protection, fair network sharing and CPU sharing to applications, we need to alleviate developer burden in IoT applications using equivalent mechanisms that provides safety guarantees and enables policy specification. Many aspects of control systems are common across and within domains -- feedback loops, multiple dependencies, safety  regulations, and mechanisms once developed can be reused across these systems. We explore methods to model such constraints and bring out challenges in enforcing these constraints.

\section{Home Heating Application}
\begin{figure}[t!]
	\includegraphics[width=\linewidth]{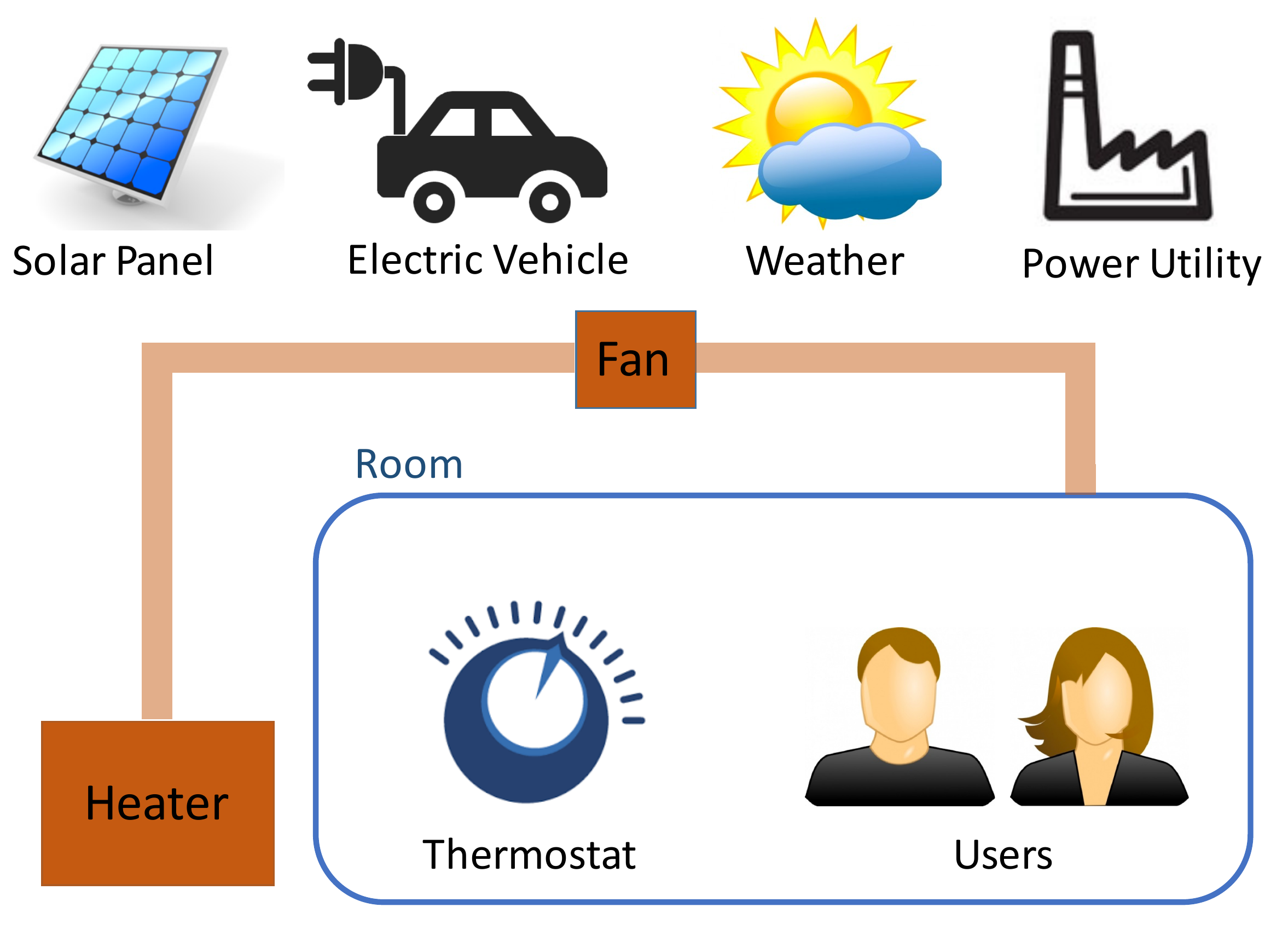}
	\caption{Home heating system with a heater for controlling air temperature and a fan for modulating air flow}
	\label{fig:home_heating_system}
\end{figure}

Consider an example of a simple home heating system to illustrate the challenges in implementing a control application. Figure \ref{fig:home_heating_system} depicts the essential components in the heating system. The heater controls the temperature of air supplied to the house and the fan controls the volume and speed of air flow. The thermostat measures the ambient temperature, and provides feedback to the heating control system. The thermostat also provides feedback to the users and captures their temperature preferences. In addition, the heating control system has external influences such as weather, solar power availability and utility events such as demand response. 

Although there are only two control knobs present in this system, the control decision has to take many constraints into consideration. User preferences dictate that temperature be within a narrow range when the house is occupied. In addition, the air supplied needs to ensure the humidity, air quality and air flow remain within a comfortable range. Equipment constraints include control within available capacity and operating points which causes minimal wear and tear. As the control system in an IoT system has information from external sources, it can optimize the operation to improve efficiency by considering outside weather conditions and availability of solar power or energy storage. Moreover, an intelligent control system needs to respond to demand response events from the utility by reducing power demand and sacrificing comfort. Thus, even the simple house heating system application needs to model constraints across many domains to make control decisions effectively. In a real system, the decision space may be further constrained by water heating system, multi-room control and user activities like cooking.  

\section{Modeling Constraints}
A first step towards reasoning about constraints would be to model them in a principled manner irrespective of domain specifics. We build upon existing data modeling methods and illustrate a few modeling strategies based on the heating system example. 

\subsection{Graph of Dependencies}
\begin{figure}[t!]
	\includegraphics[width=\linewidth]{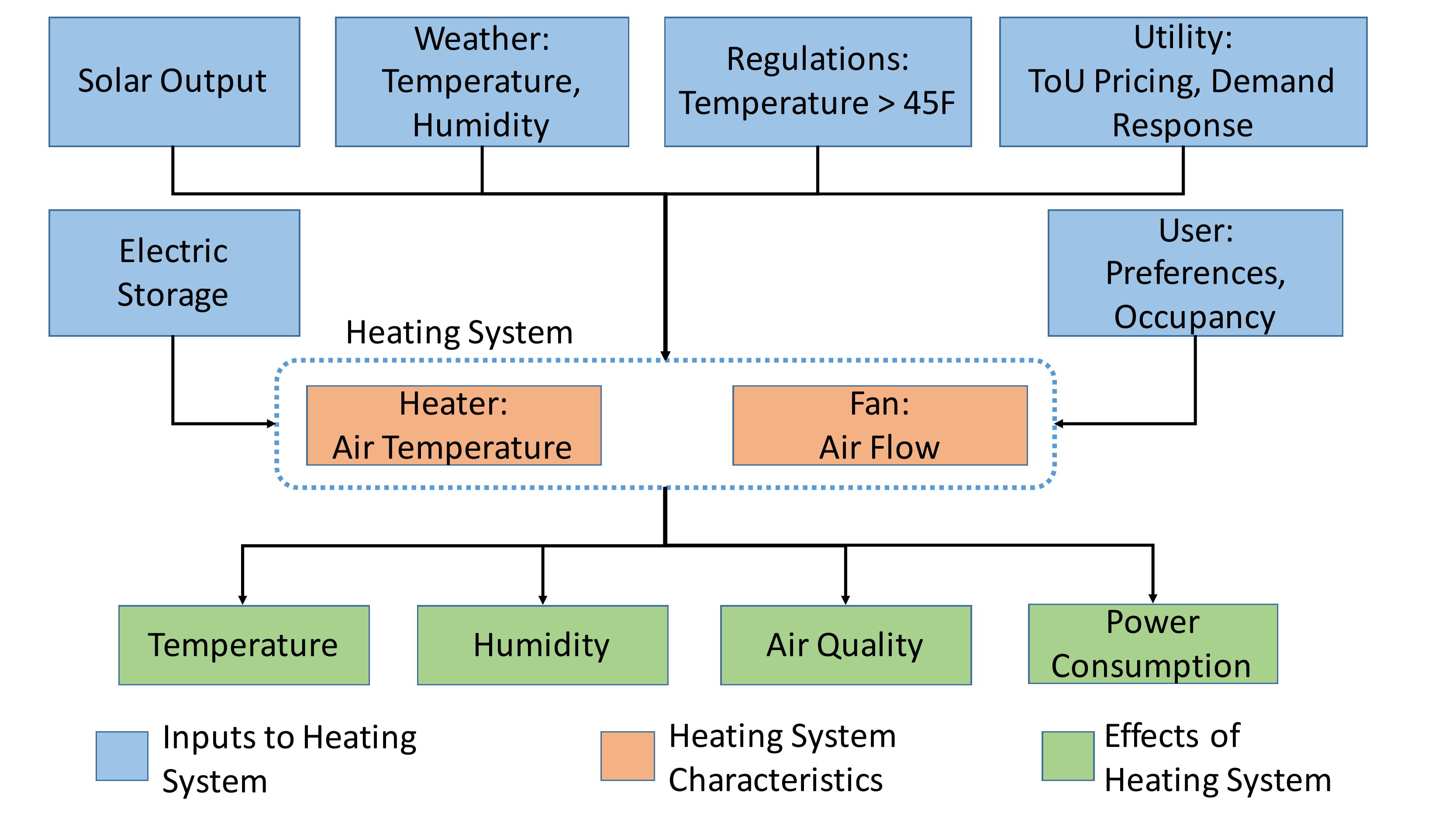}
	\caption{Model of dependencies between different modules in the home heating system}
	\label{fig:dependency_graph}
\end{figure}

Semantic ontologies provide a systematic way to model relationship between different entities of interest in a domain and has been successfully used to represent domain knowledge in a wide variety of applications~\cite{compton2012ssn}. Figure \ref{fig:dependency_graph} shows the dependencies in the home heating system. Building on this representation, we also need to quantify the relationship between two entities. For example, we need an equation representing the relationship between the outdoor temperature and the heating requirements of the house. We also need to capture the feedback loops present in the system (not shown) such as those between the temperature preferences and the measured temperature. 

\subsection{State Space Model}
\begin{figure}[t!]
	\includegraphics[width=\linewidth]{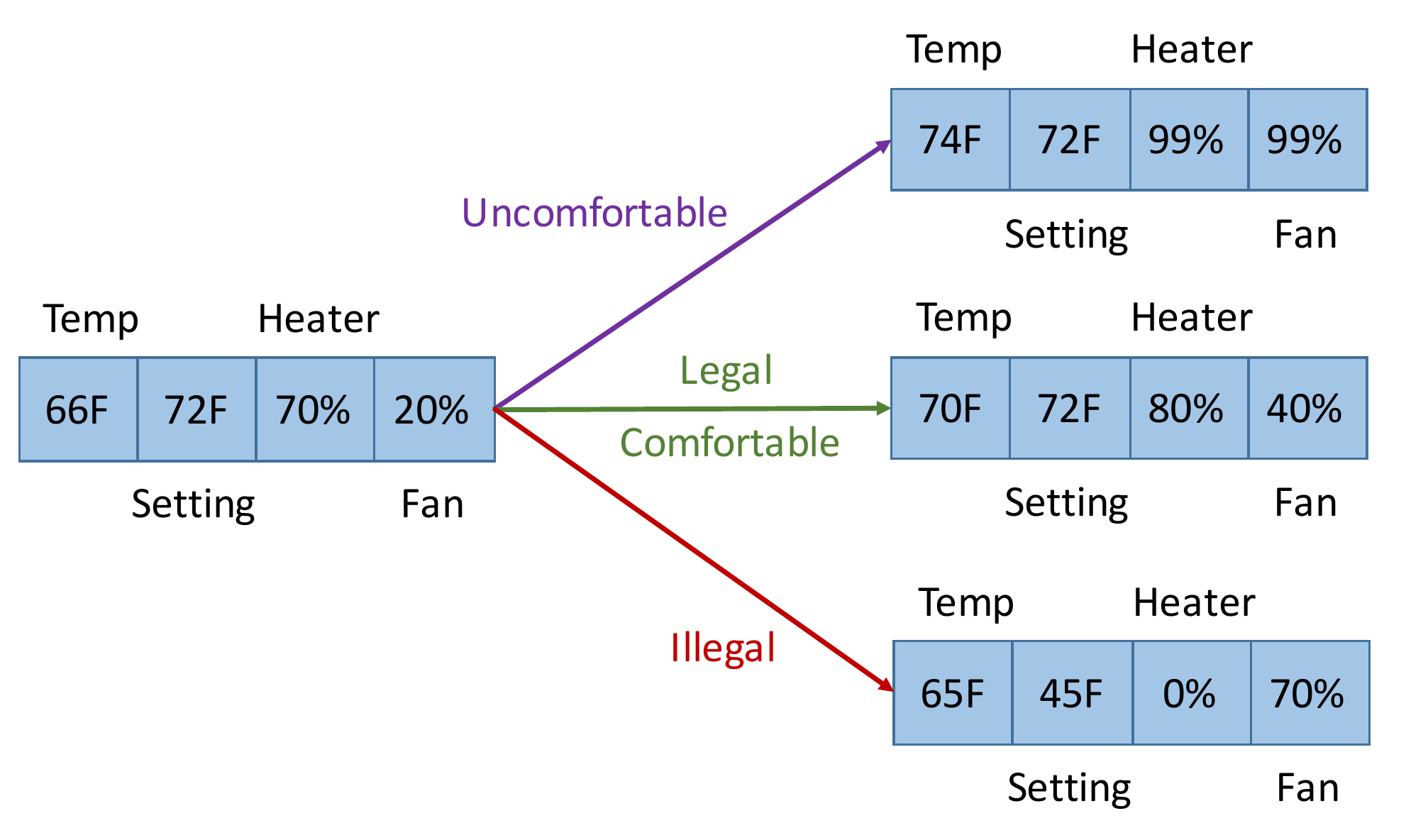}
	\caption{Example state space model capturing constraints specified by legal and health policies}
	\label{fig:state_space_model}
\end{figure}

The dependency graph provides a mapping of entities in the application domain. However, constraints in a system depend not only on the entity relationships, but also the real-time value of these entities. A model of the system state space is needed to precisely define the constraints in the system. Figure \ref{fig:state_space_model} shows a simple example of a state space model, with four essential entities in the home heating system example. The current operating point of system can be changed in multiple ways, each of which may violate the specified constraints. The state space model can represent the constraints in the system for each transition as shown in the figure.  

With such a state space model, we can analyze the system constraints (legal vs. illegal) and design tradeoffs (energy vs comfort) between different actuation decisions. A graph based representation also makes the model modular, and constraints from domains such as solar and power utility can be added as needed for each instantiation. The challenge is to represent the system state space efficiently, and reduce questions of interest to tractable problems. 

\section{Enforcing Constraints}
Modeling and enforcing constraints for IoT actuation has ramifications on many aspects of the system design. Many of the design tradeoffs need to be revisited to support actuation applications and incorporate constraints as an essential part of the system.

\subsection{Access Rights}
In modern IoT systems, access control is specified at a single sensor or actuator level~\cite{dixon2012operating}. Such an access control mechanism no longer suffices for an actuation based system. Each constraint in the system can have a large impact on the system behavior, and users/apps that specify constraints need to be access controlled. The access rights need to represent the different stakeholders in the system operation. For example, the constraints for safety policies should be specified by a health app and equipment operation constraints should be specified by a vendor app. The access control mechanism also needs to ensure actuation applications do not exceed their specified constraints, and the constraints may change with each application. For example, an emergency evacuation app will have a different set of policies governing them compared to a personalization app.

\subsection{Data Model Architecture}
The cloud provides a convenient, reliable system which can house the data models and constraint policies. The cloud also can act as a single point of contact for specifying policies across various sectors: legal, health, etc., and the constraints can be disseminated to each IoT system as the policies are updated. However, as some actuation decisions are latency sensitive, enforcement of constraints may need to be local, either using gateways or embedded actuator controllers. Thus, the communication protocol between cloud and local controllers need to be designed with care to ensure consistency and reliability.

\subsection{Policy Translation}
Low level constraints such as safe temperature settings need to be translated from high level policies. We need to develop appropriate abstractions for policy specifications so that the constraints align with expected system behavior. It is natural that conflicts may occur between policies, and mechanisms are needed for resolving conflicts both statically and dynamically. We can build upon existing application level conflict resolution strategies such as priority and manifest checks, and adopt them to constraints implementation~\cite{munir2014depsys}. We also need to provide feedback so that the developer can verify that the policies are implemented correctly. For a large system, emulation tools may be necessary to verify complex interactions between different policies.

\section{Conclusion}
Modern IoT systems are designed for data acquisition and analytics, with little support for actuation based applications. A critical part of actuation decisions is analysis of constraints in the system as dictated by laws of physics, user preferences and local laws. We explore data models that can be used for modeling constraints in an IoT system, building on top of semantic ontology models. With the example of a home heating system, we show that data models need to encapsulate the relationship between entities in an IoT system and the decision state space of the system to precisely specify constraints. Many challenges need to be addressed to enforce these constraints in an IoT system. We elucidate some of these challenges with a focus on access control mechanisms, data model architecture and policy translation.

\bibliographystyle{abbrv}
\bibliography{sigproc}  

\begin{thebibliography}{1}

\bibitem{barnaghi2012semantics}
P.~Barnaghi, W.~Wang, C.~Henson, and K.~Taylor.
\newblock {Semantics for the Internet of Things: early progress and back to the
  future}.
\newblock {\em International Journal on Semantic Web and Information Systems
  (IJSWIS)}, 8(1):1--21, 2012.

\bibitem{burgard2005coordinated}
W.~Burgard, M.~Moors, C.~Stachniss, and F.~E. Schneider.
\newblock Coordinated multi-robot exploration.
\newblock {\em Robotics, IEEE Transactions on}, 21(3):376--386, 2005.

\bibitem{compton2012ssn}
M.~Compton, P.~Barnaghi, L.~Bermudez, R.~Garc{\'\i}A-Castro, O.~Corcho, S.~Cox,
  J.~Graybeal, M.~Hauswirth, C.~Henson, A.~Herzog, et~al.
\newblock {The SSN ontology of the W3C semantic sensor network incubator
  group}.
\newblock {\em Web Semantics: Science, Services and Agents on the World Wide
  Web}, 17:25--32, 2012.

\bibitem{dawson2010smap}
S.~Dawson-Haggerty, X.~Jiang, G.~Tolle, J.~Ortiz, and D.~Culler.
\newblock {sMAP: a simple measurement and actuation profile for physical
  information}.
\newblock In {\em Proceedings of the 8th ACM Conference on Embedded Networked
  Sensor Systems}, pages 197--210. ACM, 2010.

\bibitem{dixon2012operating}
C.~Dixon, R.~Mahajan, S.~Agarwal, A.~Brush, B.~Lee, S.~Saroiu, and P.~Bahl.
\newblock An operating system for the home.
\newblock In {\em Presented as part of the 9th USENIX Symposium on Networked
  Systems Design and Implementation (NSDI 12)}, pages 337--352, 2012.

\bibitem{evans2011internet}
D.~Evans.
\newblock {The internet of things: How the next evolution of the internet is
  changing everything}.
\newblock {\em CISCO white paper}, 1:1--11, 2011.

\bibitem{munir2014depsys}
S.~Munir and J.~A. Stankovic.
\newblock Depsys: Dependency aware integration of cyber-physical systems for
  smart homes.
\newblock In {\em Cyber-Physical Systems (ICCPS), 2014 ACM/IEEE International
  Conference on}, pages 127--138. IEEE, 2014.

\bibitem{reichardt2002cartalk}
D.~Reichardt, M.~Miglietta, L.~Moretti, P.~Morsink, and W.~Schulz.
\newblock {CarTALK 2000: Safe and comfortable driving based upon
  inter-vehicle-communication}.
\newblock In {\em Intelligent Vehicle Symposium, 2002. IEEE}, volume~2, pages
  545--550. IEEE, 2002.

\end{thebibliography}

\balancecolumns 
\end{document}